\documentclass[doublecol,figures]{epl2}
\usepackage{amsmath,amssymb}
\usepackage{color}

\newcommand{\Aut}{\ensuremath{\mathrm{Aut}}}
\newcommand{\ds}{\ensuremath{{d_\mathrm{s}}}}

\title{Symmetry-based coarse-graining of evolved dynamical networks}

\author{Steffen Karalus \and Joachim Krug}

\institute{Institut f\"ur Theoretische Physik, Universit\"at zu K\"oln,
  Z\"ulpicher Str.~77, 50937~K\"oln, Germany}


\abstract{ %
  {Networks with a prescribed power-law scaling in the spectrum of the
    graph Laplacian can be generated by evolutionary optimization.  The
    Laplacian spectrum encodes the dynamical behavior of many important
    processes.  Here, the networks are evolved to exhibit subdiffusive dynamics.
    Under the additional constraint of degree-regularity, the evolved networks
    display an abundance of symmetric motifs arranged into loops and long linear
    segments.  Exploiting results from algebraic graph theory on symmetric
    networks, we find the underlying backbone structures and how they contribute
    to the spectrum.  The resulting coarse-grained networks provide an intuitive
    view of how the anomalous diffusive properties can be realized in the
    evolved structures.} }

\begin{document}

\maketitle

\section{Introduction}
Networks have become a principal tool for the modeling of complex systems in a
broad range of scientific fields~\cite{albert_statistical_2002,
  newman_structure_2003, dorogovtsev_critical_2008}.  In a dynamical network the
topology describes the couplings between individual units of a dynamical
process~\cite{barrat_dynamical_2008}.  The general underlying question of
research on dynamical networks is how the network structure shapes the global
dynamical behavior.

An important class of processes on networks are Laplacian dynamics in which the
graph Laplacian is the operator of the (linearized) time evolution.  The
Laplacian spectrum and eigenvectors then describe the overall dynamical
behavior.  This class comprises many physical processes such as dynamics of
Gaussian spring polymers~\cite{gurtovenko_generalized_2005}, transport
processes~\cite{gallos_scaling_2007}, random walks~\cite{noh_random_2004}, and
synchronization of oscillators~\cite{atay_network_2006, motter_bounding_2007,
  arenas_synchronization_2008}.

Another dynamical aspect is the evolution of network structure.  Many systems
change their connectivity structure in the course of time which affects the
global dynamical behavior.  The time scales of these two processes are often
well separated with fast dynamics and slowly responding structural evolution.
As an example, think of changes in neuronal activity in a brain on the one hand
and the formation of synaptic connections on the other hand.  In the opposite
case, when dynamics and evolution happen on similar time scales one speaks of
coevolutionary or adaptive networks~\cite{gross_adaptive_2008}.  Since the
functionality of a system is often closely associated with dynamical behavior
evolutionary forces will drive an evolving dynamical network towards some
optimized structure.

The method of network evolution adopts this strategy to find networks with
prescribed dynamical properties.  Based on rules for mutation and selection
together with a ``fitness'' function measuring the quality of a network
structure the evolution explores the configuration space towards optimal
structures.  Evolutionary optimization of networks was applied successfully in
various contexts such as modularity in changing
environments~\cite{kashtan_spontaneous_2005}, Boolean switching
dynamics~\cite{bornholdt_topological_2000, oikonomou_effects_2006,
  braunewell_reliability_2008, shao_dynamics-driven_2009, szejka_evolution_2010,
  greenbury_effect_2010}, and synchronization of
oscillators~\cite{donetti_entangled_2005, donetti_optimal_2006,
  donetti_network_2008, rad_efficient_2008, gorochowski_evolving_2010}.  It has
also been shown that networks can be reconstructed from their Laplacian spectra
by evolutionary optimization~\cite{ipsen_evolutionary_2002,
  comellas_spectral_2008}.  

{Recently, network evolution was applied to generate networks which
  display subdiffusive dynamics~\cite{karalus_network_2012}.  Here, subdiffusion
  is specified by the average return probability of a random walker that decays
  more slowly than in normal diffusion.  The return probability is directly
  related to the spectrum of the graph Laplacian via a Laplace transform.  Thus,
  in contrast to other questions like the synchronizability of oscillators where
  mainly the smallest eigenvalues contribute~\cite{arenas_synchronization_2008},
  here the entire spectrum is important. The evolution algorithm optimizes the
  spectrum with respect to a prescribed target function, which in this case is a
  pure power law characterized by the spectral dimension $\ds$.  For subdiffusive
  dynamics $\ds < 2$.}

\begin{figure}
  \centering
  \includegraphics[width=\linewidth]{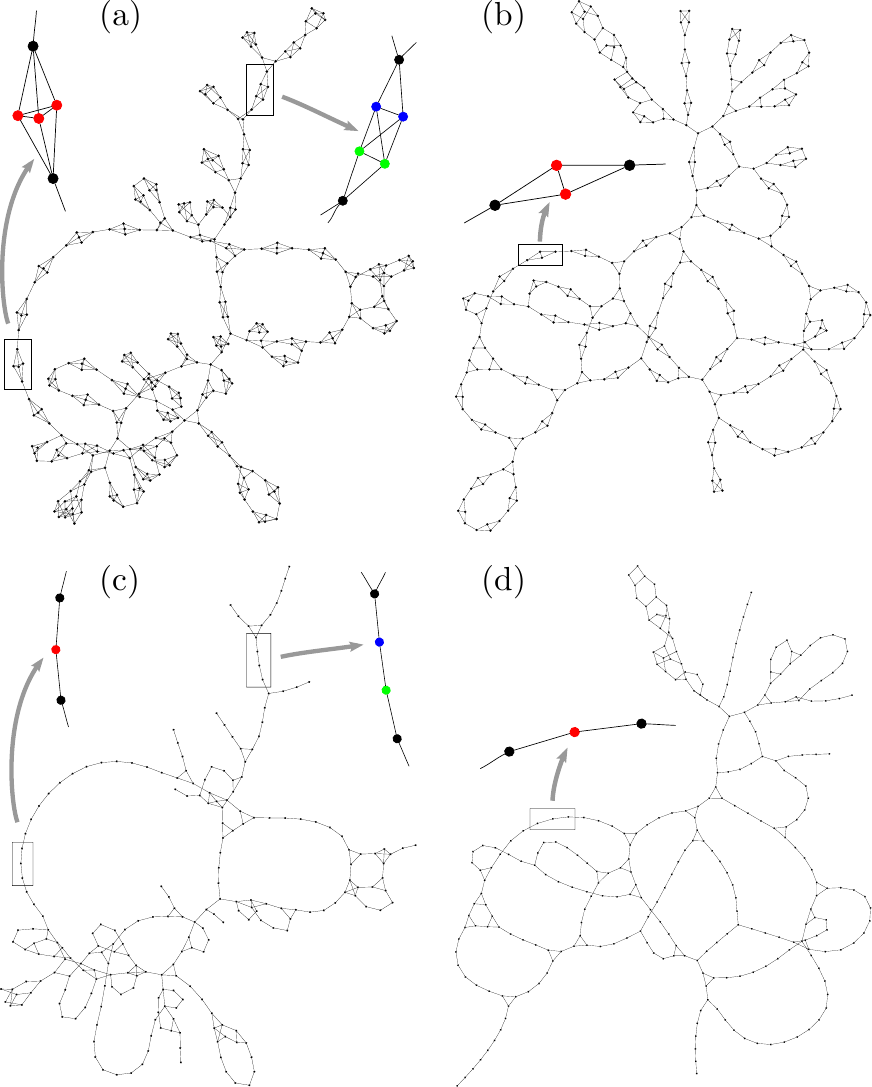}
  \caption{(Color online) {Top:} Typical $k$-regular network configurations 
    after $10^6$ evolution steps towards spectral dimension $\ds = 1.4$ starting from
    {(a)}~a square lattice with $k=4$, $N=361${, $M=722$} 
    and {(b)}~a honeycomb
    lattice with $k=3$, $N=360${, $M=540$}.
    {Bottom: The corresponding s-quotients which are of sizes
      (c)~$N=230$, $M=294$ and (d)~$N=285$, $M=337$.  In the enlarged segments,
      the colors indicate the orbits to which the respective vertices belong.}}
  \label{fig:net_final}
\end{figure}
To elucidate how the spectral properties are encoded in the structure of the
evolved networks, in the present work we apply the method of
ref.~\cite{karalus_network_2012} to \textit{regular} networks, i.e., networks in
which each vertex has the same number of neighbors. Two typical network
configurations evolved towards a target spectral dimension of $\ds = 1.4$ are
shown in fig.~\ref{fig:net_final}{(a) and (b)}.  They exhibit {an} interesting,
modular structure consisting of small symmetric motifs arranged into
fractal-like loops and linear segments of different lengths on larger scales.
{Given that subdiffusive transport is common in comb-like and
  self-similar structures~\cite{havlin_diffusion_2002}, it is very tempting to
  attribute the subdiffusive character of the networks to this modular
  structure. We know, however, that primarily spectral and not structural
  properties determine the dynamical behavior~\cite{atay_network_2006,
    samukhin_laplacian_2008} and that the spectrum of a network can, in general,
  not be constructed from the spectra of its subnetworks.  Nevertheless, in the
  present case the symmetry of the networks implies that this can be done in a
  systematic way.}

In the following we exploit concepts of algebraic graph theory to extract the
large-scale structure of the evolved networks.  This is achieved by identifying
the symmetries associated with the small-scale motifs and constructing the
corresponding quotient network.  The quotient graph and in particular the
corresponding simple graph represent a systematic coarse-graining of the
original networks, and their spectra are considerably closer to the target
function of the evolutionary algorithm.  To set the stage for this
investigation, we begin by providing some background on spectral graph theory
and network symmetries.

\section{Spectra of dynamical networks}
A formal description of a network with $N$ vertices is given by the adjacency
matrix $A$.  It is an $N \times N$ matrix with elements $A_{ij} = 1$ if vertex
$i$ is connected with vertex $j$ and $A_{ij} = 0$ otherwise.  For an undirected
network $A$ is symmetric, $A_{ij} = A_{ji}$.  In the case of a directed network
one has to specify whether $A_{ij}$ describes the presence of the an edge from
$i$ to $j$ or vice versa.  Both conventions exist in the literature.  The graph
Laplacian $L$ is another matrix describing the structure of an undirected simple
network (no self-loops, no multi-edges).  It is defined as $L = D - A$ where $D$
is the diagonal matrix of vertex degrees $k_i$, $D_{ij} = k_i \delta_{ij}$.  The
degree $k_i = \sum_j A_{ij}$ of a vertex~$i$ is the number edges incident to the
vertex.  Note that this operator is sometimes called the algebraic Laplacian to
distinguish it from similar operators such as different versions of the
normalized Laplacian.

Diffusion processes form a very important class of dynamics.  They describe,
e.g., how a substance spreads in a medium or an opinion in a society.  To be
more specific, we consider a process where the flux of some substance along each
edge in the network is proportional to the difference in the amount at the end
vertices.  The time evolution operator of this process on an undirected network
is the graph Laplacian.  An important global characteristic of such a process is
the average probability~$P_0(t)$ for a random walker to return to its starting
point at time~$t$.  
{$P_0(t)$ is the Laplace transform of the Laplacian eigenvalue
  density~$\rho(\lambda)$, $P_0(t) = \int_0^\infty \mathrm{d} \lambda
  \exp(-\lambda t) \rho(\lambda)$~\cite{havlin_diffusion_2002,
    samukhin_laplacian_2008}.  Hence,} if the integrated eigenvalue density
{$I(\lambda) = \int \mathrm{d} \lambda \, \rho(\lambda)$} scales as a
power law, $I(\lambda) \propto \lambda^{\ds / 2}$,
then $P_0(t)$ decays
as a power law, $P_0(t) \propto t^{-\ds/2}$.  {The power-law exponent~$\ds$ is
  the spectral dimension of the network.  For regular lattices, it coincides
  with the Euclidean dimension~\cite{burioni_random_2005}.}

Since the quotient graph to be constructed below is directed, we need to
generalize the graph Laplacian to directed networks.  For this purpose we
consider a diffusion process on a directed network, possibly containing
multi-edges and self-loops.  Let $x_i$ describe the amount of the diffusing
quantity on vertex $i$.  The flux of this quantity along each outgoing edge
should be proportional to $x_i$.  Then $x_i$ decreases at rate $c x_i$ for each
outgoing edge $i \to j$, and increases at rate $c x_j$ for each incoming edge $j
\to i$ where $c$ is {a diffusion coefficient}.
Following the convention that $A_{ij}$ is given by the number of directed edges
from $i$ to $j$ this yields the set of coupled differential equations
\begin{align}
  \dot{x}_i
  &= \sum_j A_{ji} c x_j - \sum_j A_{ij} c x_i
  = c \sum_j A_{ji} x_j - c x_i k_i^\mathrm{out} \nonumber\\
  &= c \sum_j \left( A_{ji} - \delta_{ji} k_j^\mathrm{out} \right) x_j
  = -c \sum_j L_{ji}^\mathrm{out} x_j
  \label{eq:diffusion}
\end{align}
where $k_i^\mathrm{out} = \sum_j A_{ij}$ denotes the out-degree of vertex $i$ and 
\begin{equation}
  L_{ji}^\mathrm{out} = k_j^\mathrm{out} \delta_{ji} - A_{ji}
  \label{eq:outLap}
\end{equation}
are the elements of the out-degree Laplacian.  So, when considering diffusion
processes on a network the out-degree Laplacian is the correct generalization of
the graph Laplacian as time evolution operator to directed networks.

\section{Symmetries in networks}
{In a symmetric network, the spectrum and, correspondingly, the set of
  eigenvectors can be split into two classes.  Some of the eigenpairs stem from
  the symmetric motifs and are inherited by the network as a whole while the
  remaining ones are the eigenpairs of the underlying backbone structure, the
  quotient network.  In order to make this statement more precise and formally
  tractable we first summarize some basic notions}
on the mathematical description of network symmetry.  More details{,
  exact definitions,} and proofs can be found in
ref.~\cite{macarthur_spectral_2009, godsil_algebraic_2004}.

The symmetry of a network $G$ is described by its automorphism group $\Aut(G)$.
An \emph{automorphism} is a permutation of vertices which does not change the
network structure (preserves the adjacency).  A network is called
\emph{symmetric} if it has a non-trivial automorphism group.  By a direct
product decomposition $\Aut(G) = H_1 \times H_2 \times \ldots \times H_k$ the
automorphism group can be split into \emph{geometric factors} $H_i$ which act
independently on the network.  A \emph{symmetric motif} $\mathcal{M}_H$ of a
network is the induced subgraph on the support of a geometric factor $H$.
Intuitively speaking, a symmetric motif $\mathcal{M}_H$ is a minimal subgraph
whose vertices are moved by $H$.

The action of a network's automorphism group $\Aut(G)$ partitions the vertex set
$V$ into disjoint equivalence classes called \emph{orbits}.  The orbit of a
vertex $v \in V$ is the set of vertices $\Delta(v) \subset V$ to which $v$ maps
under the action of $\Aut(G)$,
\begin{equation}
  \Delta(v) = \{ gv \in V \, | \, g \in \Aut(G) \} \,.
\end{equation}
{In the enlarged parts of fig.~\ref{fig:net_final}(a) and (b) examples
  of how the vertices in a symmetric motif are grouped into orbits are indicated
  by their colors.  In general,} it may{, however,} be necessary to
swap a larger fraction of a network in order to map two vertices in the same
orbit onto each other without altering the adjacency.  Nevertheless all vertices
in the same orbit are structurally equivalent (perform the same function in the
network).

The \emph{quotient graph} is a network coarse-graining based on this structural
equivalence.  The set of vertices of the quotient $Q = G / \mathbf{\Delta}$ is
given by the orbits $\mathbf{\Delta} = \{ \Delta_1 , \Delta_2 , \ldots ,
\Delta_r \}$ of the network $G$.  Note that due to the structural equivalence
the number of neighbors in $\Delta_y$ of a vertex $v \in \Delta_x$, called
$q_{xy}$, does not depend on the choice of $v$ but only on $x$ and $y$.  A
partition with this property is called \emph{equitable} which has important
consequences for the relation between the spectra of $G$ and $Q$ (see below).
The quotient graph $Q$ of a network $G$ has vertex set $\mathbf{\Delta}$ and
adjacency matrix $\{q_{xy}\}_{x,y = 1, \ldots, r}$.
It is by definition directed and generally contains multi-edges and self-loops.
If one is working with simple networks this may be inconvenient.  As a further
simplification the \emph{s-quotient}~$Q_\mathrm{s}$ is defined as the underlying
simple graph of the quotient.  It already captures essential properties such as
the size and adjacency of the quotient.  {Fig.~\ref{fig:net_final}(c) and (d)
  show the s-quotients of the two networks above.  The colored vertices in the
  enlarged segments indicate the corresponding orbits in the parent networks.}

A key result from algebraic graph theory~\cite{godsil_algebraic_2004} relates
the adjacency spectra and eigenvectors of a network $G$ and its quotient graph
$Q$ (or, in general, any quotient $G/\pi$ from an equitable partition $\pi$).
All eigenvalues of the quotient are also eigenvalues of the parent network.
Moreover, to each eigenpair $(\lambda,\mathbf{v}=(v_1,\ldots,v_m))$ of $Q$ there
exists an eigenpair $(\lambda,\mathbf{\bar{v}}=(\bar{v}_1,\ldots,\bar{v}_n))$ of
$G$ with $\bar{v}_i = v_x$ for all $i \in \Delta_x$ (eigenvectors are constant
$v_x$ on each orbit $\Delta_x$).  These eigenvectors are called \emph{lifted}
from the eigenvectors of the quotient.  The remaining eigenpairs of $G$ are
constructed from the redundant eigenpairs of the symmetric motifs of $G$: If
$(\lambda,\mathbf{v}=(v_1,\ldots,v_k))$ is an eigenpair of the symmetric motif
$\mathcal{M}$---isolated from the rest of the network---which is
\emph{redundant} (i.e.\ $\sum_{i \in \Delta} v_i = 0$ for each orbit $\Delta \in
\mathcal{M}$) then $(\lambda,\mathbf{\hat{v}})$ is an eigenpair of $G$ with
$\hat{v}_i = v_i$ for all $i \in \mathcal{M}$ and $\hat{v}_i = 0$ for $i \notin
\mathcal{M}$ (eigenvectors are localized on the symmetric motifs).

To summarize, the adjacency eigenpairs of a symmetric network $G$ fall into two
classes.  (1) The eigenpairs of the quotient of $G$ are also eigenpairs of $G$
with constant values of the eigenvectors on the orbits.  (2) The redundant
eigenpairs of the symmetric motifs of $G$ are also eigenpairs of $G$ with
eigenvectors localized on the respective motif.

All these statements for the adjacency spectrum and eigenvectors are elaborated
and proven in ref.~\cite{macarthur_spectral_2009}.  In the appendix we show that
they also hold for the Laplacian spectrum and eigenvectors.

For practical computations of automorphism groups efficient and easily
applicable algorithms are available such as the \texttt{nauty}
program~\cite{mckay_practical_2014} which was used in this study to find orbits
and construct network quotients.

\section{Application to regular evolved networks}
In ref.~\cite{karalus_network_2012} the method of network evolution was applied
to construct networks with a given non-trivial spectral dimension
$d_\mathrm{s}$.  The algorithm searches the configuration space of connected
networks with a given size for network structures which resemble the power law
in the Laplacian spectrum as closely as possible.  The introduced distance
measure $\mathcal{D}$ (called $\Delta$ in ref.~\cite{karalus_network_2012}) is
defined as the integral over the squared difference between integrated
eigenvalue densities.  In contrast to other proposed spectral
distances~\cite{jurman_introduction_2011} it can be applied to compare the
spectrum with a general function as well as with other spectra.  However, since
$\mathcal{D}$ inherently depends on the number of vertices, networks of
different sizes cannot be compared directly.  The most simple rules for mutation
and selection were applied, namely a random movement of a single edge and the
acceptance of a proposed move if $\mathcal{D}$ decreases and the whole network
remains connected.  Starting from regular lattices and random graphs networks
were successfully evolved under the constraints of a fixed number of vertices
and a fixed average degree.

An even stricter condition is to fix the individual degree of each vertex, as
for example in a $k$-regular network where all vertices have the same degree
$k$.  Is it---under this additional constraint---still possible to construct
networks with a given non-trivial value of the spectral
dimension~$d_\mathrm{s}$?  To answer this question we applied the network
evolution algorithm starting from square lattices ($k=4$) and honeycomb lattices
($k=3$) and corresponding $4$- and $3$-regular random networks.  As in the
{preceding} study~\cite{karalus_network_2012} it turns out that the
evolution does not depend on the two choices (lattice or random network) of
initial conditions.  The conservation of the degree sequence (or the individual
degree of each node) in the course of the evolution can be realized by an
``edge-crossing'' update.  In each evolution step two edges $v_1$--$v_2$ and
$v_3$--$v_4$ are randomly chosen and ``crossed'' to $v_1$--$v_4$ and
$v_3$--$v_2$.  When the edges are drawn one has to make sure that the four
vertices $v_1,v_2,v_3,v_4$ are all distinct and that the edges $v_1$--$v_4$ and
$v_3$--$v_2$ do not exist before the update.

\begin{figure}
  \centering
  \includegraphics[width=\linewidth]{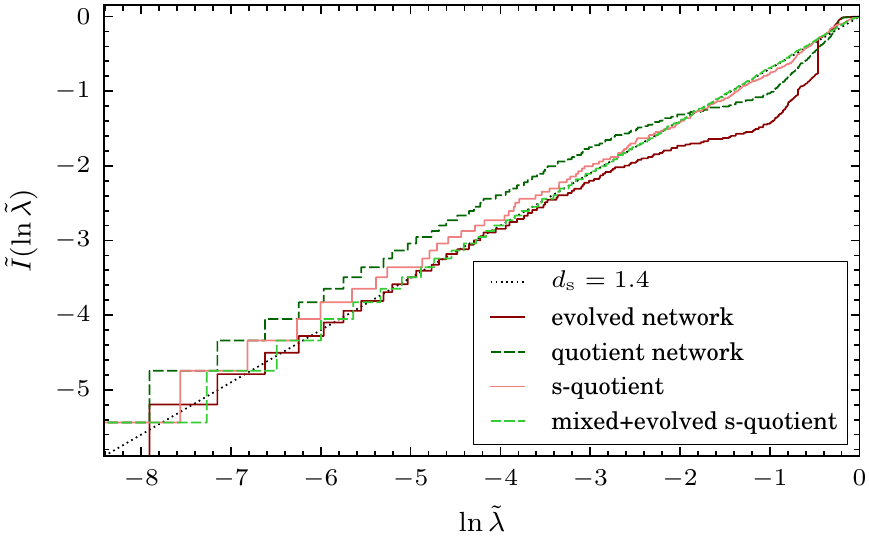}
  \caption{(Color online) Logarithmically integrated Laplacian eigenvalue
    densities of the 4-regular evolved network from
    fig.~\ref{fig:net_final}{(a)} and its quotient, s-quotient
    (fig.~{\ref{fig:net_final}(c)})
    and an evolved network of the same size (number of vertices and number of
    edges) as the s-quotient. The eigenvalue density is plotted as a function of
    $\ln \tilde \lambda$, where $\tilde \lambda = \lambda/\lambda_\mathrm{max}
    \leq 1$ is the eigenvalue normalized by the largest eigenvalue of the
    network \cite{karalus_network_2012}. The black dotted line indicates the
    target integrated density with $\ds = 1.4$.}
  \label{fig:intdens}
\end{figure}

The evolutions were run for $10^6$ time steps towards a target value of the
spectral dimension of $d_\mathrm{s} = 1.4$.  Typical network configurations at
the end of the evolutions of $3$- and $4$-regular networks are shown in
fig.~\ref{fig:net_final}{(a) and (b)}.  These networks appear to have
similar large-scale structures of long loops built up by very frequently
appearing symmetric motifs on small scales.
\begin{figure}
  \centering
  \begin{minipage}[c]{.49\linewidth}
    \includegraphics[width=\linewidth]{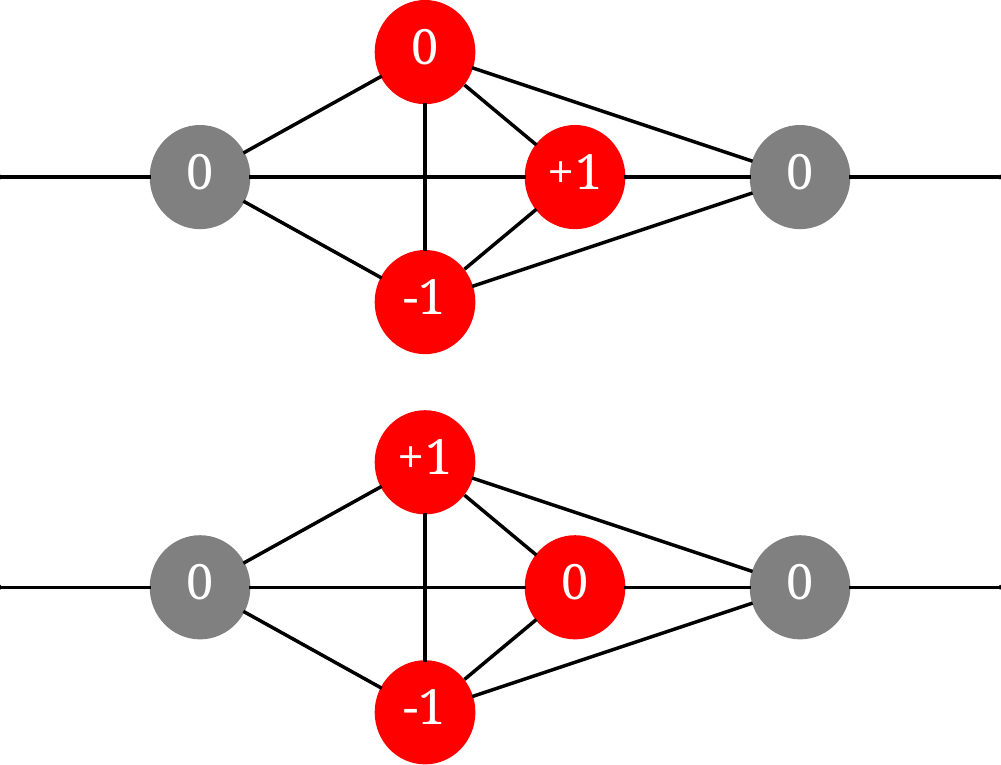}
  \end{minipage}
  \hfill
  \begin{minipage}[c]{.49\linewidth}
    \includegraphics[width=\linewidth]{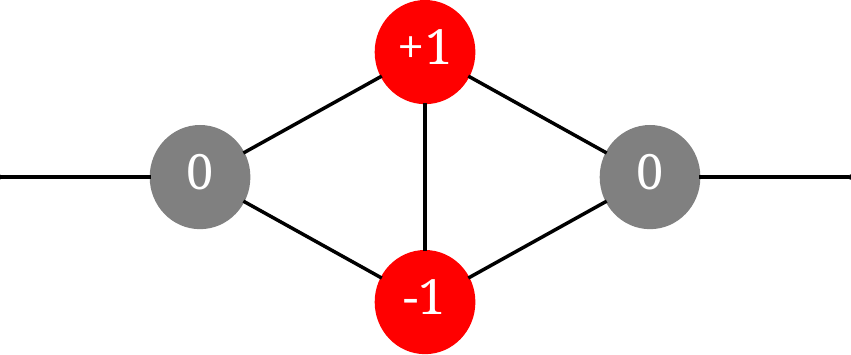}
  \end{minipage}
  \caption{(Color online) Redundant Laplacian eigenvectors of the most prominent
    motifs in 4-regular (left) and 3-regular (right) evolved networks as
    depicted in fig.~\ref{fig:net_final}{(a) and (b)}.  The symmetric
    vertices are colored red.  The corresponding eigenvalues are $\{5,5\}$
    (left) and $\{4\}$ (right).}
  \label{fig:motifs}
\end{figure}
The most abundant of these motifs are shown in fig.~\ref{fig:motifs}.  They
consist of a complete subnetwork of $k-1$ vertices, each of which is joined to
two boundary vertices connecting to the rest of the network.  The vertices are
labeled by their redundant Laplacian eigenvectors and the corresponding
eigenvalues are $\{5,5\}$ in the 4-regular and $\{4\}$ in the 3-regular case.
These are also eigenvalues of the network as a whole, which explains the high
degeneracy giving rise to the large step in the integrated eigenvalue density in
fig.~\ref{fig:intdens}.  The motifs are assembled to form large-scale structures
with long loops which appear to be quite similar in both cases.
The large-scale structures of these networks become more evident in their
s-quotients depicted in fig.~{\ref{fig:net_final}(c) and (d)}.
One can clearly see the arrangement of loops of different lengths.  Seen as
a coarse-graining transformation, the s-quotient on average reduces the network
sizes from $N=361$ vertices and $M=722$ edges to $N \approx 229(6)$ and $M
\approx 302(17)$ for 4-regular parent networks and from $N=360$, $M=540$ to $N
\approx 281(6)$, $M \approx 330(9)$ for 3-regular parent networks (standard
deviations in parentheses).

How to assess the quality of the s-quotients with respect to the optimization
goal of the evolution?  The ultimate goal of the evolution algorithm is to
construct networks with a Laplacian spectrum which resembles the power law
$I(\lambda) \propto \lambda^{d_\mathrm{s}/2}$.  Figure \ref{fig:intdens}
displays the spectra---as logarithmically integrated densities $\tilde{I}(\ln
\tilde{\lambda})$ (see eq.~(5) of ref.~\cite{karalus_network_2012} for the exact
definition)---for an exemplary 4-regular evolved network, its quotient,
s-quotient, and an evolved network of the same size as the s-quotient (``mixed
and evolved s-quotient'').  The latter class of networks was analyzed in order
to get an idea of how the quotients and s-quotients compare to unconstrained
evolved networks of the same size.  They are constructed by optimizing networks
of the same size (same numbers of vertices and edges) by means of the
unrestricted version of the evolution algorithm (conserving these numbers and
the connectedness only).  Technically, since the s-quotients are very sparse
networks, instead of generating random networks, the edges of the s-quotients
are randomized (keeping the whole network connected) and then the evolution is
run.

As predicted by theory, the spectrum of the quotient resembles the spectrum of
its parent up to the highly degenerate redundant eigenvalues of the symmetric
motifs. By removing the corresponding steps in the integrated eigenvalue density
the fit to the target spectrum is improved in the regime of large eigenvalues,
whereas the region of small $\lambda$ is unaffected. In contrast, there is no
rigorous relation between the spectra of the parent network and the s-quotient.
Therefore it was not to be expected that the spectrum of the s-quotient
resembles the power law even more closely, as seen in the figure.  In terms of
the average distance measure $\bar{\mathcal{D}}$ estimated from $100$
realizations of the evolution, the quotients have a value of $\bar{\mathcal{D}}
= 0.09(1)$ which is reduced by a factor of three in the s-quotients
($\bar{\mathcal{D}} = 0.03(2)$).  Nevertheless, the unconstrained evolution of
networks of the same size succeeds in finding networks with a spectrum even
closer to the desired power law, reaching an average value of $\bar{\mathcal{D}}
= 0.0045(1)$.  However, similar to the networks obtained in
ref.~\cite{karalus_network_2012}, the networks resulting from the unconstrained
evolution algorithm cannot be easily spread out and visualized in a plane.

\begin{figure}
  \centering
  \includegraphics[width=.8\linewidth]{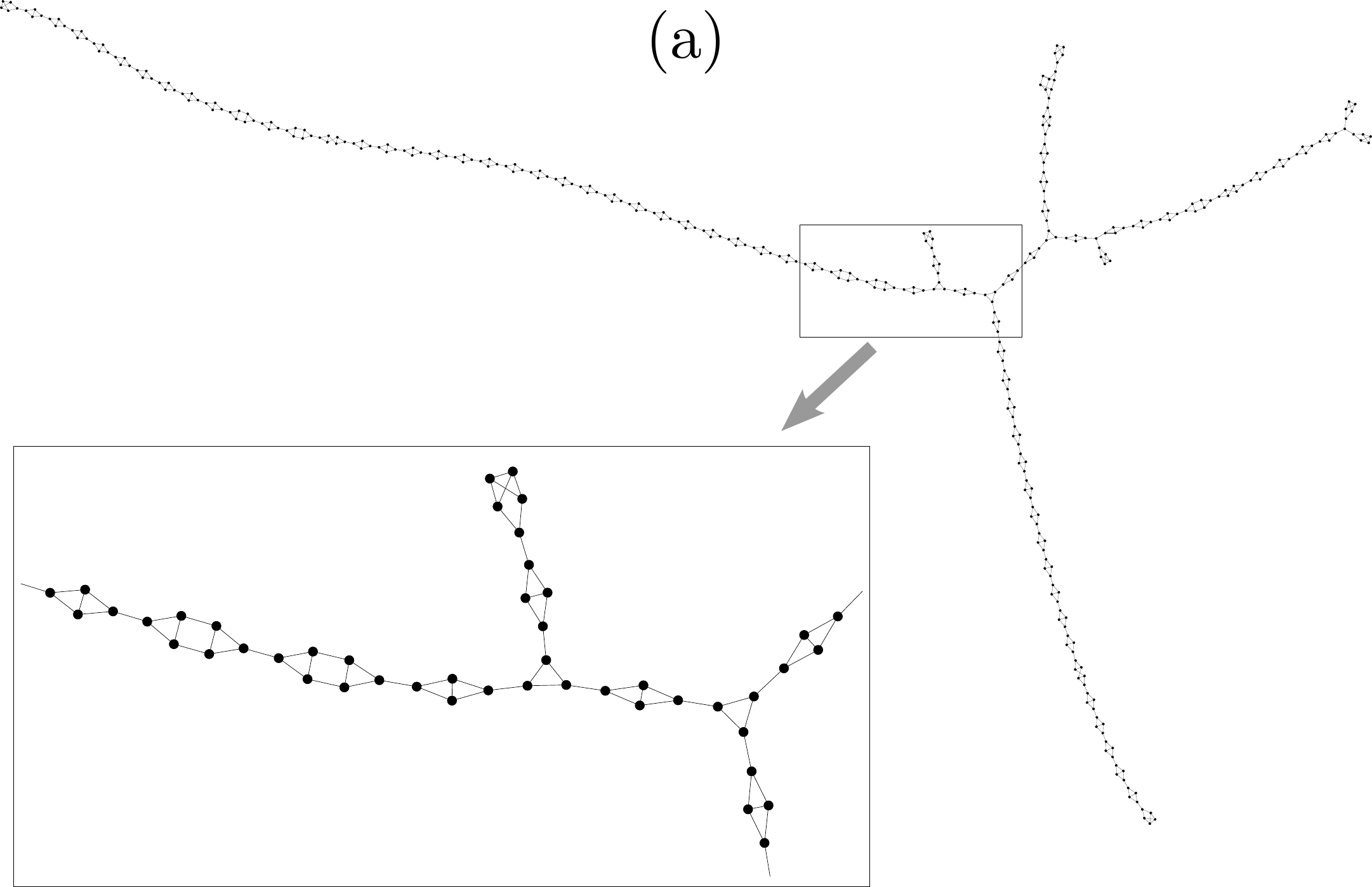}\\
  \vspace{2mm}
  \includegraphics[width=.8\linewidth]{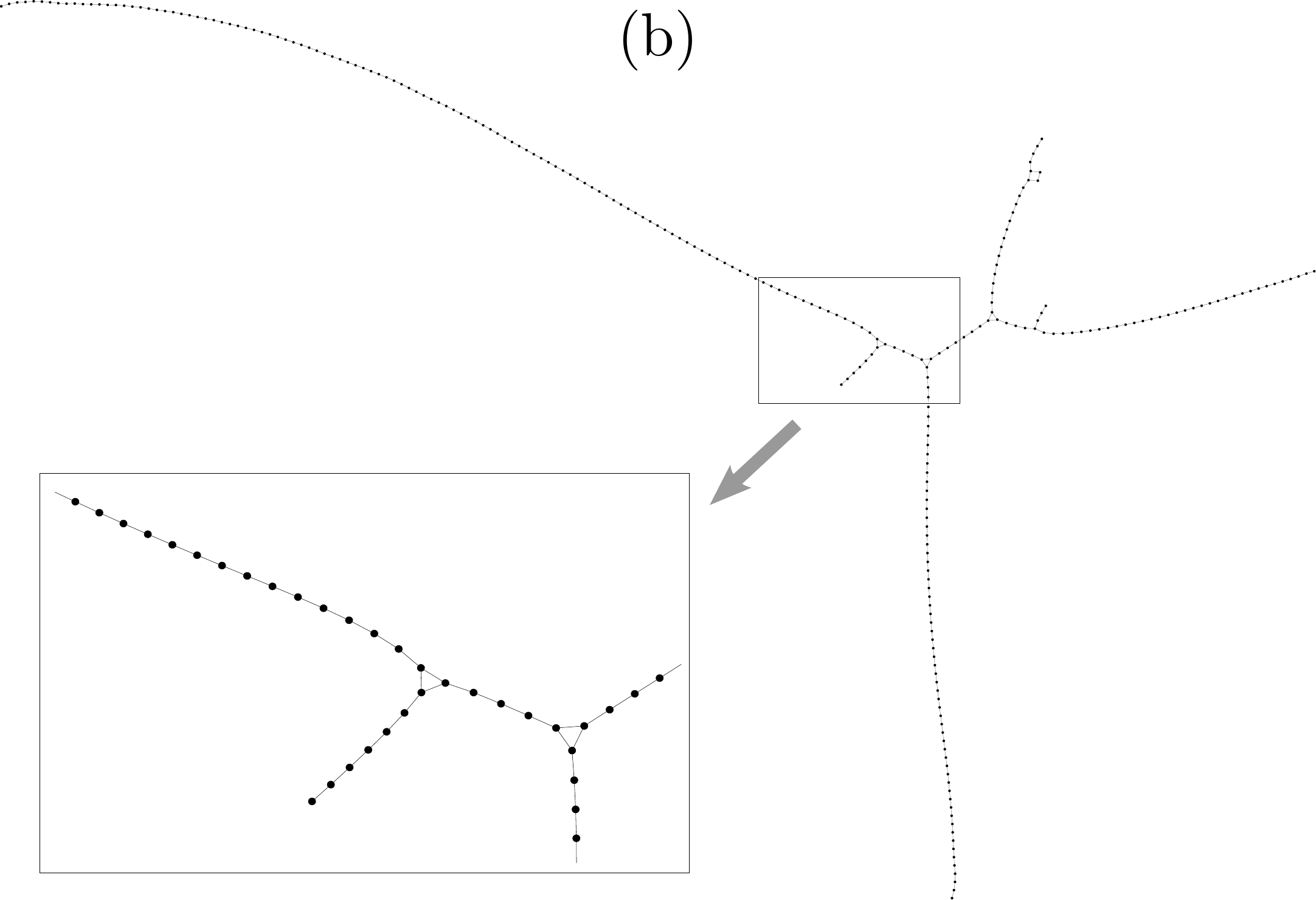}
  \caption{(a) Typical $3$-regular network configuration after $10^6$ evolution
    steps towards $\ds = 1.1$ starting from a honeycomb lattice with $N=360$,
    $M=540$ and (b) corresponding s-quotient, $N=260$, $M=263$.}
  \label{fig:ds11}
\end{figure}
What happens for a more extreme evolution target?  Since the evolution runs were
started from two-dimensional lattices it should be more difficult to find
networks with a spectral dimension closer to $1$.  Fig.~\ref{fig:ds11} shows a
typical network configuration from an evolution towards $\ds = 1.1$ together
with the 
{corresponding} s-quotient.  In this case, the basic motifs are
arranged to form even longer linear chains instead of loops resulting in an
almost one-dimensional s-quotient.

An important question is \textit{why} the evolved networks assume the observed
shape of symmetric motifs assembled in loops and linear chains (see
fig.~\ref{fig:net_final}{(a) and (b)} and \ref{fig:ds11}(a)).  A
possible explanation is that they have too many edges to form structures with a
low spectral dimension.  The formation of the symmetric motifs makes it possible
to ``hide'' the excess edges in the motifs and form large-scale structures with
the desired scaling in the small Laplacian eigenvalues.  Additionally, the
motifs may effectively act as particle traps and thus slow down the diffusion
process.  The formation of loops and linear chains on larger scales is brought
out more clearly in the coarse-grained s-quotients in
fig.~{\ref{fig:net_final}(c) and (d)}
and \ref{fig:ds11}(b) where the symmetric motifs are removed.
\begin{figure}
  \centering
  \includegraphics[width=\linewidth]{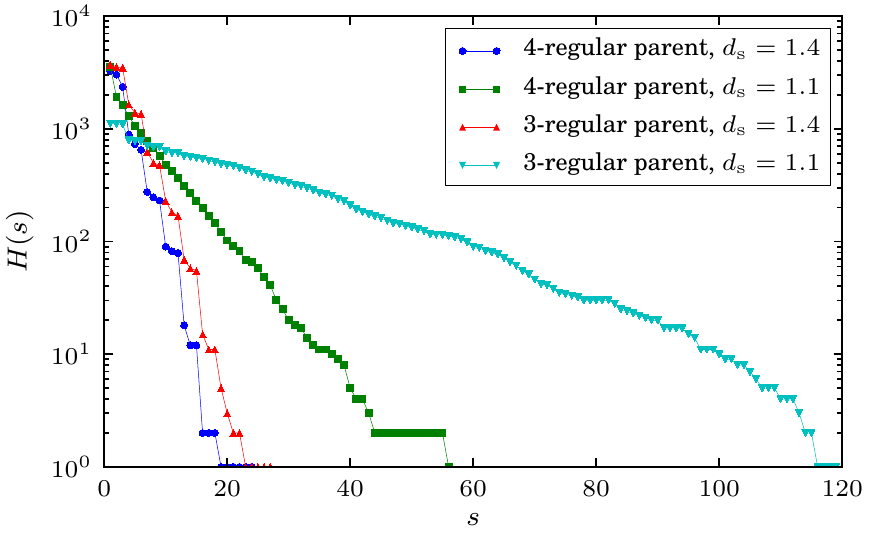}
  \caption{(Color online) Histogram of linear segment lengths $s$ in s-quotients
    from 4-regular and 3-regular evolved networks for different target spectral
    dimensions $\ds$.  Shown are cumulative histograms $H(s) = \int_s^\infty
    h(t) \,\mathrm{d}t$ of the absolute frequencies $h(s)$ for $100$
    realizations of each.}
  \label{fig:linSeg}
\end{figure}
As a first approach towards quantifying the loop structure we analyzed
the lengths of linear
segments (connected maximal subgraphs of nodes with a maximum degree two) in the
s-quotients in fig.~\ref{fig:linSeg}.  
We observe a very broad distribution in this quantity.  However, the networks
appear to be too small for a systematic scaling analysis.

\section{Conclusions}
We presented and analyzed subdiffusive networks found by evolutionary
optimization.  Subdiffusion is known to be observed in various network
structures such as regular fractals, percolation clusters or combs with
power-law distributed teeth lengths~\cite{havlin_diffusion_2002}.  In the latter,
the delay is caused by the time a random walker spends on the teeth of the comb.
In the networks presented here, we observe backbone structures with loops and
dangling ends of different length scales which seem to be the main source of
subdiffusive behavior.  The networks are, however, too small for a quantitative
analysis of the distribution of these lengths.  A more in-depth study on this
question would be an interesting continuation of this work.

\acknowledgements
We thank \textsc{Thilo Gross} for sharing his knowledge on network symmetries,
\textsc{Markus Porto} for his support in early stages of this work, and
\textsc{Sungmin Hwang} for valuable comments and discussions.  We gratefully
acknowledge partial funding by the \textit{Stu\-dien\-stif\-tung des deut\-schen
  Vol\-kes} and the \textit{Bonn-Cologne Graduate School of Physics and
  Astronomy}.

\section{Appendix}
We show that the relations between adjacency spectra and eigenvectors of a
network~$G$ on the one hand and its quotient $Q$ and symmetric motifs on the
other hand also hold for the Laplacian spectrum and eigenvectors.  The reasoning
is the same as in Appendix A of ref.~\cite{macarthur_spectral_2009}.  First,
recall that the characteristic matrix~$P$ of the equitable partition $\pi = \{
C_1 , \ldots , C_r \}$ is defined by $P_{ix} = 1$ if vertex $i \in C_x$ and
$P_{ix} = 0$ otherwise. It remains to prove that the space $W$ spanned by the
column vectors of
$P$ is also $L$-invariant.  For this, $W$ is $L$-invariant ($Lu \in W$ for all
$u \in W$) if and only if there exists a matrix~$\tilde{L}$ such that
\begin{equation}
  L P = P \tilde L \,.
\end{equation}
$\tilde L$ is then the graph Laplacian of the quotient network $Q = G / \pi$.
Let $B$ ($B_{xy}=q_{xy}$ for the orbit partition above) be the adjacency
matrix of the quotient.  Define $\tilde L = \tilde D - B$ with $\tilde{D}_{xy} =
\tilde{k}_x \delta_{xy}$ and $\tilde{k}_x = \sum_y B_{xy}$.  Since $AP=PB$ it
remains to show that $DP=P\tilde{D}$.  The left hand side reads
\begin{equation}
  (DP)_{ix} = \sum_j D_{ij} P_{jx} = \sum_j k_i \delta_{ij} P_{jx} = k_i P_{ix}
  \label{eq:lhs}
\end{equation}
which is the degree of vertex $i$ if $i \in C_x$ and zero otherwise.  The right
hand side reads
\begin{align}
  (P \tilde D)_{ix}
  &= \sum_y P_{iy} \tilde D_{yx}
  = \sum_y P_{iy} \tilde k_y \delta_{yx} \nonumber\\
  &= P_{ix} \tilde k_x = P_{ix} \sum_y B_{xy} \,.
  \label{eq:rhs}
\end{align}
Since $B_{xy}$ is the number of neighbors in $C_y$ of a vertex from $C_x$,
$\sum_y B_{xy}$ gives the total number of neighbors of a vertex in $C_x$ and
eq.~\eqref{eq:rhs} evaluates to the degree of vertex~$i$ if it belongs to $C_x$
and zero otherwise.  This is the same as eq.~\eqref{eq:lhs} which completes
the proof.

Note that the definition of the Laplacian $\tilde L$ of the quotient graph is
consistent with the out-degree Laplacian~$L^\mathrm{out}$ defined in
eq.~\eqref{eq:outLap}.  In particular, the definition of $q_{xy}$ corresponds 
to the convention that $A_{ij}$ describes directed edges from $i$ to $j$.


\end{document}